\documentclass{article}
\usepackage{psfig}
\usepackage{emulateapj}

\begin{document}

%%% For double spaced apj style (submitted form): aasms4
%usepackage{aasms4}
%%% for single space apj style preprint form:     aaspp4
%\usepackage{aaspp4}
%%% For single space apj style with two columns: aas2pp4
%\usepackage{aas2pp4}
%%% For an Apj feel (warning, bugs) : emulateapj}
%%% For two column apj style:  aaspptwo
%\usepackage{aaspptwo}

\newcommand{\gtorder}{\mathrel{\raise.3ex\hbox{$>$}\mkern-14mu
             \lower0.6ex\hbox{$\sim$}}}
\newcommand{\ltorder}{\mathrel{\raise.3ex\hbox{$<$}\mkern-14mu
             \lower0.6ex\hbox{$\sim$}}}
\newcommand{\proptwid}{\mathrel{\raise.3ex\hbox{$\propto$}\mkern-14mu
             \lower0.6ex\hbox{$\sim$}}}

\newcommand{\pa}{\partial}
\newcommand{\na}{\nabla}
\newcommand{\ti}{\times}

\slugcomment{To Appear in The Astrophysical Journal; Feb. 20, 1999; Volume 512}

\lefthead{Begelman}
\righthead{A Model for the Moving `Wisps' in the Crab Nebula}

\title{A MODEL FOR THE MOVING `WISPS' IN THE CRAB NEBULA}

\author{Mitchell C. Begelman\altaffilmark{1}}
\affil{JILA, University of Colorado and National Institute of Standards and Technology \\
Campus Box 440, Boulder, CO 80309-0440; mitch@jila.colorado.edu}

\altaffiltext{1}{Also Department of Astrophysical and Planetary Sciences, University of
Colorado, Boulder}

\begin{abstract}
  I propose that the moving `wisps' near the center of the Crab Nebula
  result from nonlinear Kelvin-Helmholtz instabilities in the equatorial
  plane of the shocked pulsar wind. Recent observations suggest that the
  wisps trace out circular wavefronts in this plane, expanding radially at
  speeds $ \ltorder c /3$. Instabilities could develop if there is
  sufficient velocity shear between a faster-moving equatorial zone and a
  slower moving shocked pulsar wind at higher latitudes. The development of
  shear could be related to the existence of a neutral sheet --- with weak
  magnetic field --- in the equatorial zone, and could also be related to a
  recent suggestion by Begelman that the magnetic field in the Crab pulsar
  wind is much stronger than had been thought. I show that plausible
  conditions could lead to the growth of instabilities at the radii and
  speeds observed, and that their nonlinear development could lead to the
  appearance of sharp wisplike features.
\end{abstract}

\keywords{radiation mechanisms: instabilities --- ISM: individual (Crab
  Nebula) --- ISM: jets and outflows --- pulsars: individual: Crab pulsar
  --- supernova remnants}

\section{Introduction}

Recent observations have clarified the geometry and kinematics of the
bright optical arcs, commonly known as `wisps,' that are situated $\approx
7-15'' $ northwest of the Crab pulsar.  The variability of these features
has been known for decades (Lampland 1921; Oort \& Walraven 1956; Scargle
1969), but as recently as 1995 it was still unclear whether they fluctuated
in position and/or intensity, or exhibited coherent motion (Hester et
al.~1995).  Systematic monitoring over a period of 3.5 years by Tanvir,
Thomson, \& Tsikarishvili (1997) now seems to establish that the wisps move
radially outward from the pulsar, with speeds $\ltorder c/3$ (assuming a
distance to the Crab of 2 kpc).  Moreover, the wisps appear to represent
approaching segments of circular wavefronts confined to the equatorial
plane of the pulsar, which is tipped $\sim 30^\circ $ to the line of sight.
High resolution images of the wisps (Hester et al.~1995), obtained with the
WFPC2 on board {\it Hubble Space Telescope} show that the wisps are
extremely narrow ($\sim 0.''2$ in width), with synchrotron surface
brightnesses exceeding the immediate background by an order of magnitude or
more.

A few models in the literature have attempted to explain the physical
nature of the wisps.  A large fraction of the Crab pulsar's spindown energy
is thought to emerge via a relativistic, magnetohydrodynamic wind, which
suffers a strong shock in the vicinity of the wisps (Rees \& Gunn 1974;
Kennel \& Coroniti 1984). The detailed model by Gallant \& Arons (1994)
associates the multiple wisps with internal structure in the pulsar wind
termination shock. Other models interpret the wisps as propagating
magnetosonic (Woltjer 1958; Barnes \& Scargle 1973) or drift (Chedia et
al.~1997) waves. Lou (1996, 1998) regards the wisps as originating in the
ultrarelativistic wind upstream of the shock.  Hester et al.~(1995) propose
that the wisps are instabilities driven by synchrotron cooling in the flow.
Other modelers have regarded the wisps as generically associated with the
wind termination shock, at various levels of detail (e.g., Rees \& Gunn
1974; Kundt \& Krotscheck 1980; Kennel \& Coroniti 1984). These models have
varying degrees of success in explaining the systematic outward motion
found by Tanvir et al.~(1997).

In this paper I present a new model that can account for the wisps'
systematic outward velocity, their apparent confinement to a thin
equatorial sheet in the pulsar wind, and their sharp brightness profiles.
I propose that a narrow equatorial zone of the shocked pulsar wind is
moving somewhat faster than the wind at higher latitudes.  The thin
equatorial zone can be identified physically with the ``neutral sheet"
across which the direction of the toroidal magnetic field reverses. Such a
zone must be present in any pulsar wind.  In \S~4 I will discuss physical
effects that may give rise to the shear, but for the present I will simply
assume that it exists. The shear between the zones gives rise to
Kelvin-Helmholtz (K-H) instabilities which cause the equatorial sheet to
ripple. If the sheet is radiating and the ripples grow to sufficiently
nonlinear amplitudes, then the surface brightness distribution of the sheet
will exhibit narrow, bright arcs that can be identified as the wisps.  In
\S~2 I discuss the K-H instability as it may apply in the shocked pulsar
wind, and show that it can lead to ripples with roughly the observed
spacings and pattern speeds.  In \S~3 I show how our line of sight through
a sufficiently rippled, emitting sheet can give rise to sharp arc-like
features on the near and far sides of the sheet (with the leading side
favored due to the Doppler effect), regardless of whether additional
dissipative effects operate in the sheet.  Finally, in \S~4 I discuss
possible reasons for the development of shear and observational
implications of the model.

\eject

\section{KELVIN-HELMHOLTZ INSTABILITY OF THE SHEAR ZONE}

For the purpose of a stability analysis, we model the equilibrium state of
the equatorial shear zone as a uniform plane-parallel slab of unmagnetized
relativistic fluid, with thickness $2 H$ in the $z-$direction, moving with
velocity $u c \hat x$ with respect to a stationary background medium.  The
exterior fluid also has a relativistic equation of state, as well as
containing a uniform magnetic field oriented perpendicular to the flow
velocity, $B_{\rm ex} \hat y$.  The assumption of a negligible magnetic
field in the equatorial zone is motivated by its identification with the
neutral sheet in the pulsar wind. We parameterize the field strength in the
high-latitude region by its beta-parameter with respect to the external gas
pressure, $\beta \equiv 8\pi P_{\rm ex} / B_{\rm ex}^2 $.  Pressure balance
in the $z-$direction then requires the pressure in the unmagnetized slab to
satisfy $P_{\rm in} = P_{\rm ex} + B_{\rm ex}^2/8\pi $. Note that the
particle densities in both fluids are irrelevant since the inertia is
dominated by the relativistic energy density.

The Kelvin-Helmholtz (K-H) instability for a slab geometry with various
physical conditions has been discussed extensively in the literature (e.g.,
Gill 1965; Ferrari, Massaglia, \& Trussoni 1982; Payne \& Cohn 1985; Hardee
\& Norman 1988; Hardee et al.~1992).  Since the plane-parallel model is
intended to approximate axisymmetric radial flow, we consider only modes
propagating parallel to the flow velocity. Given perturbations of the form
$f(z) \exp[ i ( \omega t - k x) ] $, it is straightforward to derive the
dispersion relation for the specified conditions.  Defining the
dimensionless quantities $w\equiv \omega H / c$, $\nu \equiv k c / \omega$,
and
\begin{eqnarray}
\kappa_1^2 & = & { 3 (1 - \nu u)^2 - (\nu - u)^2   \over 1 - u^2 }, \\
\kappa_2^2 & = & \nu ^2 - { 3 (1 + 2 \beta) \over (3 + 2 \beta) } , \\
{\cal A} & = & {(1- u^2)(2\beta + 1)\over 2 (\beta +1) (1 - \nu u)^2  \kappa_2}  ,  
\end{eqnarray}
we obtain 
\begin{equation}  
\tan(w \kappa_1 ) = - {\cal A} \kappa_1   
\end{equation}
for the modes that are antisymmetric (AS) about $z = 0$ (kink modes), and
$\cot(w \kappa_1) = {\cal A} \kappa_1 $ for the symmetric (S, or pinch)
modes.  The form of the dispersion relation given above is analogous to
that introduced by Gill (1965) and used by many subsequent authors.  For
example, to relate our variables to the nonrelativistic magnetohydrodynamic
limit studied by Hardee et al.~(1992), we make the following
identifications with their parameters $\beta_{\rm in}$, $\beta_{\rm ex}$,
$\chi_{\rm in}$, $\chi_{\rm ex}$: $\kappa_1 \rightarrow \beta_{\rm in} H /
w$, $\kappa_2 \rightarrow i \beta_{\rm ex} H / w$, ${\cal A} \rightarrow -
(\chi_{\rm ex}H)/ ( \chi_{\rm in} w \kappa_2 ) $. To satisfy boundary
conditions at $|z| \rightarrow \infty$, the real part of $w \kappa_2$ must
be positive.  Note that special relativistic effects are included in the
derivation.

The mode structure predicted by eqs.~(1)--(4) is qualitatively similar to
that found in nonrelativistic treatments of slab jets, both with and
without magnetic fields (Hardee \& Norman 1988; Hardee et al.~1992).  We
are interested in the spatial growth of modes with real driving
frequencies, so we set $\omega$ (and $w$) real and $k = k_R + i k_I$; $k_I
> 0$ then corresponds to a growing mode.  In addition to the fundamental AS
and S modes which are unstable at all flow velocities, internal reflection
modes appear when the flow is sufficiently supersonic (and
supermagnetosonic; see Hardee et al.~1992). However, these latter modes
have phase and group velocities that exceed the effective sound speeds in
both the unmagnetized ($c/\sqrt{3}$) and magnetized ($> c/\sqrt{3}$)
regions.  If Tanvir et al.~(1997) are correct that the wisps move outward
at speeds $\ltorder c/ 3$, then the pattern speed is highly {\it subsonic}.

We will assume (based on the appearance of the wisps; see \S~3) that we are
dealing with the fundamental AS mode in the long-wavelength limit, $|k H |
\ll 1$. We can then approximate the dispersion relation by ${\cal A} + w
\approx 0$. These modes correspond to side-to-side displacements ---
ripples --- of the slab. Since the $e$-folding length of the instability,
$k_I^{-1}$, is a monotonically decreasing function of $\omega$, all
wavelengths shorter than a certain value will have reached nonlinearity at
any given radius in the wind. We will associate the wisps with the {\it
  longest} wavelength (i.e., lowest frequency) modes that have room to
grow, since these are expected to dominate the large-scale structure of the
ripples. Thus, the dominant modes have $k_I \ \sim$ few $\times r^{-1}$,
where $r$ is the radius at which the ripples appear.

Phase velocities $v_p = \omega / k_R c \equiv \nu_R^{-1}$ of
very-long-wavelength modes vary as $v_p \proptwid (kH)^{1/2} \proptwid
\omega^{1/3}$, i.e., they decrease with increasing wavelength and
decreasing frequency. The observed pattern speed of a weakly unstable
linear wave packet is the group velocity, $\partial \omega / \partial k_R$,
but the physical significance of this quantity is ill-defined for modes
with $k_I \sim k_R$, as is the case here.  Moreover, we are considering
modes that have grown to nonlinear amplitudes, by which point the pattern
speeds may have changed.  In their numerical simulations of nonrelativistic
slab jets, Norman \& Hardee (1988) find that the linear phase velocity
gives a more accurate estimate of the nonlinear pattern speed than does the
group velocity, and we will adopt this assumption here. Therefore, if the
high-latitude wind has decelerated to $\ll c/3$ at the point at which we
see the wisps, then the equatorial wind would have to be traveling with
speed $u c$ significantly faster than $c/3$ (from a relativistic point of
view) in order to obtain the observed pattern speeds of $\sim c/ 3$. Note
that the standard spherical shocked wind model for the Crab Nebula (Rees \&
Gunn 1974) predicts decreasing subsonic velocities $\sim (c/3) (r/
r_s)^{-2}$ outside the shock radius, $r_s$.  (The exact radial dependence
is sensitive to the two-dimensional flow pattern.) Even if the equatorial
wind speed significantly exceeded $c/3$, it would not necessarily have to
be highly supersonic (i.e., $> c / \sqrt{3} \sim 0.58 c$).  Figure 1 shows
$k_R H$ (solid lines) and $k_I H$ (dashed lines) as a function of slab
velocity $u$, for negligible ($\beta \rightarrow \infty$: heavy lines) and
equipartition ($\beta = 1$: light lines) magnetic fields, with the phase
velocity fixed at 1/3. The magnetic field strength has little effect on the
curves.  Note also that $kH$ is already quite small ($\ltorder 0.15$) for
transsonic flow ($u = 0.58$), and decreases steadily for higher flow
velocities.

Setting $\nu_R^{-1}= 1/3$ corresponds to the case in which the outflow
speed of the high-latitude wind can be neglected.  If the high-latitude
wind were still moving outward at a sizable fraction of $c/3$ in the wisp
region, then similar net pattern speeds could be attained with a slower
equatorial wind. In this case, it would be even easier to obtain very small
values of $|kH|$, which are desirable for producing sharp, intense wisp
features (see \S~3). Note, however, that the limiting case of an equatorial
wind moving only slightly faster than the high-latitude wind is not
conducive to producing wisps with the observed spacing through K-H
instability, since the modes would be advected outward before they had time
to grow.

\begin{figure*}
\centerline{
\psfig{figure=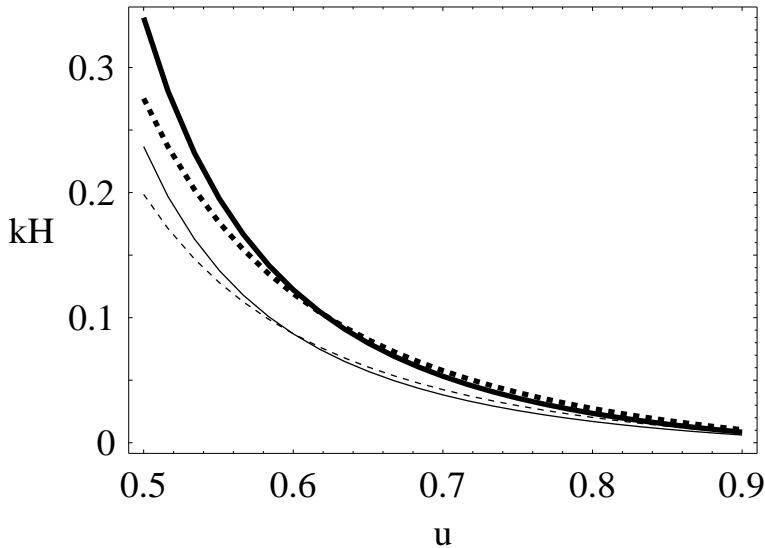,width=0.55\textwidth} }
\bigskip
\caption {Real (solid) and imaginary (dashed) parts of $kH$ as a
  function of the slab speed $u c$, assuming a phase velocity $\omega /
  k_R$ fixed at $c/ 3$. Heavy and light curves correspond to negligible and
  equipartition values of the ambient magnetic field, respectively. }
\label{fig:CrabKH1}
\end{figure*}

At the end of the epoch studied by Tanvir et al.~(1997), there were two
bright wisps roughly $7''$ and $15''$ from the pulsar.  If we argue that
the wavelength $2\pi / k_R$ of the dominant wisp mode is approximately
$r/2$ and that this corresponds to $\sim 6$ growth lengths (consistent with
fact that $k_I \sim k_R$), then we have $k_R r \sim 4\pi$. From the above
estimates based on the theory of K-H instability, we expect $k_R H \ltorder
0.1$ for $u \gtorder 0.6$, implying that the modes potentially have aspect
ratios $r/8H \gtorder 16$. The ratio can be much larger if the equatorial
wind is faster than $0.6 c$ or if the speed of the high-latitude wind is
taken into account.  These values are in the range required to obtain the
observed narrowness and brightness contrasts of the wisps from viewing
angle effects, as we shall now describe.
     
\section{NARROW, BRIGHT FEATURES FROM A RIPPLED SHEET}

Two of the most striking features of the wisps are their narrowness and
their brightness contrast with respect to the local background. Assuming
that the wisps are cylindrical filaments, Hester et al.~(1995) estimated
their equipartition (i.e., minimum) pressures at 28--54 times the mean
equipartition pressure in the nebula ($\langle P_{\rm eq} \rangle \sim
7\times 10^{-9}$ dyne cm$^{-2}$: Trimble 1982). If the wisps were highly
overpressured with respect to the {\it local} medium, then they would have
to be out of dynamical equilibrium with their surroundings, i.e., they
would have to be shocks. In our observing frame these putative strong
shocks are traveling {\it downstream} at $\sim c/ 3$, but in the fluid
frame they would have to be traveling {\it upstream} at somewhat more than
the sound speed, $c/ \sqrt{3}$. Applying the relativistic velocity addition
law, this would imply that the equatorial wind would have to be traveling
outward at highly supersonic speed, $\gtorder 0.8 c$.  This is difficult to
reconcile with the hypothesis that the wisps exist in the postshock region
where the wind is subsonic or, at best, mildly supersonic.

Even if the equatorial wind is highly supersonic in the wisp region, it is
hard to imagine what would create a set of shocks that always move outward
at $\sim c/3$.  Suppose the high-latitude wind went through a shock at
$r_s$, but the equatorial wind punched through and remained supersonic.
The sudden jump in pressure would drive oblique shocks into the equatorial
wind as it crossed $r_s$, with the obliquity adjusted so that the shock
fronts track the position of $r_s$.  Thus, these shocks would not move
systematically outward unless $r_s$ also moved systematically outward,
which seems unlikely.  Similarly, internal shocks caused by intrinsic
fluctuations in the wind should exhibit both inward and outward motions.

Fortunately, the high surface brightnesses can be explained without
resorting to large overpressures.  First, the mean pressure in the Crab may
well be several times larger than the equipartition pressure.  Moreover,
the local pressure in the center of the nebula may be several times the
nebular mean, due to magnetic tension effects (Kennel \& Coroniti 1984;
Begelman \& Li 1992). A simple theory of the wind termination shock gives
the pressure just downstream from the shock as $P_s \sim 9\times 10^{-9}
(L_w / 5 \times 10^{38} \ {\rm erg \ s}^{-1} ) (r_s / 0.1 \ {\rm pc} )^{-2}
$ dyne cm$^{-2}$, where $L_w$ is the wind power. Recall that we are
associating the wisps with K-H modes that grow downstream from the shock.
Since the inner wisp is situated at $\sim 0.1$ pc, the actual radius of the
shock might be 2--3 times smaller, suggesting that the ambient pressure in
the wisp region could be as much as $\sim 4-10$ times larger than $\langle
P_{\rm eq} \rangle $.  Second, Doppler boosting will cause one to
overestimate the equipartition pressure in the approaching wisps by a
factor $\sim {\cal D}^{1.5}$, where ${\cal D}$ is the Doppler factor (Heinz
\& Begelman 1997).  For wind speeds $\gtorder 0.5 c$ and viewing angles
$\approx 30^\circ$, this leads to a further decrease in the intrinsic
equipartition pressure by a factor $\sim 2$.
 
The remaining factor $\sim 3-6$ in inferred overpressure can be explained
away if we interpret the wisps not as cylindrical filaments but as thin
radiating sheets, viewed nearly edge-on. This geometry can arise naturally
from the nonlinear development of antisymmetric Kelvin-Helmholtz modes in a
thin equatorial wind, provided that $k_R H$ is sufficiently small and the
amplitude of the ripples grows sufficiently large.  If the equatorial wind
has uniform emissivity and the depth of our line of sight through the wind
is $2 \eta H \gg H$, then the inferred equipartition pressure goes down by
a factor $\sim \eta^{-1/2}$.  Thus, a depth-to-width ratio of $\eta \sim
10-30$ is probably sufficient to explain the observed surface brightness in
the wisps. These ratios are consistent with the predictions of the
Kelvin-Helmholtz instability, provided that the viewing geometry is
favorable.

This simple model can account not only for the brightness enhancement in
the wisps, but also for their narrowness and their apparent confinement to
a pair of arcs along the near and far sides of the wavefront.  To
illustrate this, we consider a rippled axisymmetric slab with upper and
lower surfaces (at some fixed time) given by
\begin{equation}
z'_\pm = H (\pm 1 + a \sin kr') 
\end{equation}
where $r' = (x'^2 + y'^2)^{1/2}$ in terms of Cartesian coordinates defined
in the equatorial plane of the slab.  Now suppose we observe the slab along
a line of sight in the $x'-z'$ plane, at an angle $\theta \approx 30^\circ$
from the slab's equator. We define observer's coordinates $(x,y,z)$
according to $x' = x \csc \theta - z \cos \theta$, $y' = y$, $z'=z \sin
\theta$, where $(x, y)$ lie in the sky-plane and $z$ denotes position along
the line of sight.  Our line of sight enters and exits the slab at
\begin{equation}
z_\pm = H \csc \theta \left[ \pm 1 + a \sin \left(kr_0 - {kx\cot\theta \over r_0} z_\pm  \right) \right]  ,  
\end{equation}
where $r_0 = (x^2 \csc^2 \theta + y^2 )^{1/2} $ and we have neglected terms
of second- and higher order in $z_\pm / r_0$.

If the slab radiates uniformly, its brightness is proportional to the
length of the line of sight between the entrance and exit points, $(z_+ -
z_-)$. Since the unrippled slab ($a = 0$) has uniform brightness $2 H \csc
\theta$, we can quantify the enhancement due to rippling through the
multiplicative factor
\begin{equation}
\Delta \equiv  (z_+ - z_-) \sin \theta / 2H .  
\end{equation}
Defining the quantities $ s \equiv (z_+ + z_-) \sin \theta / 2H $ and $q
\equiv kH x \cos\theta / (r_0 \sin^2\theta)$, we can determine $\Delta$
from the following pair of equations:
\begin{eqnarray}
\Delta & = & 1 - a \sin(q \Delta) \cos (kr_0 - qs) \\
 s & = & a \cos(q \Delta) \sin(kr_0 - qs) . 
\end{eqnarray}
For fixed $a$ and $q$ (corresponding to a fixed position angle on the sky,
relative to the center of the slab), maximum values of $\Delta$ occur for
$s = 0$ and $\sin kr_0 = 0$, and satisfy $\Delta_m = 1 \pm a \sin
(q\Delta_m)$. For a given value of $\Delta_m$, there are an infinite number
of contours in the $a - q $ plane which have identical shapes but are
displaced from one another according to $q_n(a) = q_0(a) +
n(\pi/\Delta_m)$, where $n$ is an integer. Figure 2 shows a set of contours
corresponding to $q_0$, which is the relevant branch if the wisps have
simple brightness contours with a single maximum at $y=0$.  For ripples of
fixed amplitude $a$, the maximum possible enhancement, $\Delta = a +1$,
occurs for $\cos (qa + q) = 0$.

\begin{figure*}
\centerline{
\psfig{figure=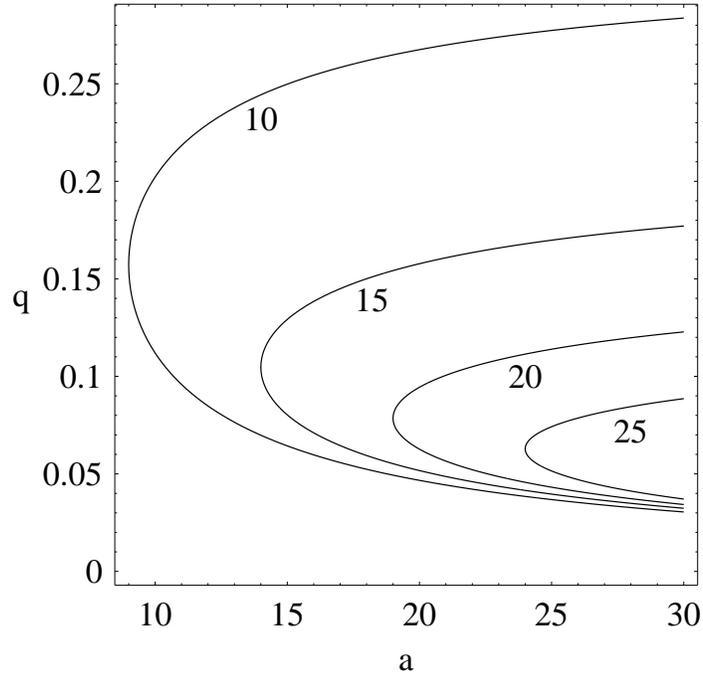,width=0.5\textwidth} }
\vskip -0.25 true in
\caption {Contours of maximum brightness enhancement due to viewing
  geometry through a uniformly emitting rippled slab, in the $a-q$ plane.
  $a$ is the amplitude of the ripples in units of slab thickness $H$ and
  $q$ is the product of scaled wavenumber $kH$ with geometric factors.
  Contours are labeled with the value of the brightness enhancement factor,
  $\Delta_m$.}
\label{fig:CrabKH2}
\end{figure*}

From the definition of $q$, we see that the maximum value of $q$ for fixed
$x$ occurs at $y = 0$. Thus, if the wisps are to be brightest on the
leading and trailing sides of the wavefront (i.e., at $y=0$), we must have
$\partial \Delta_m / \partial q |_a > 0$.  For the $q_0$ branch, this
corresponds to the condition $kH \cot\theta < \pi /(2 a + 2)$. Figure 3
shows a representative surface brightness plot for this case.

\begin{figure*}
\centerline{
\psfig{figure=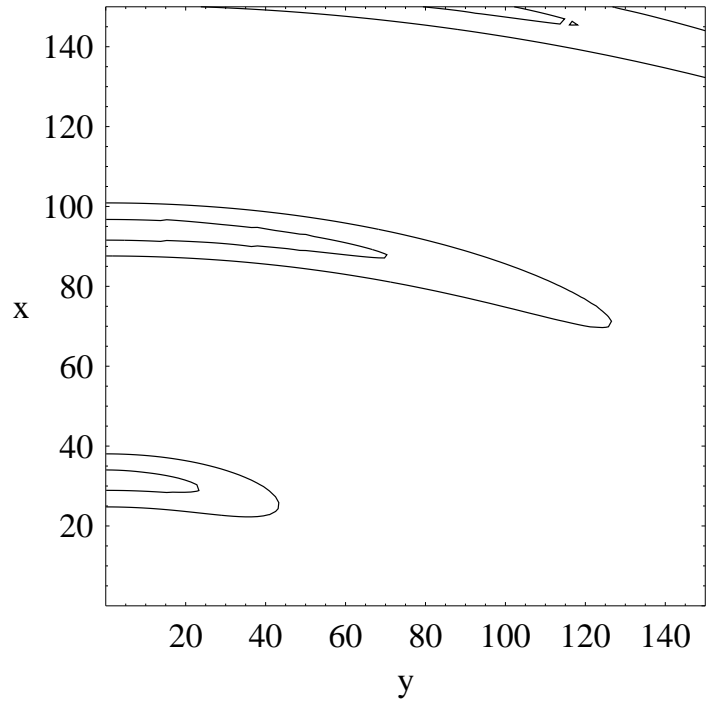,width=0.5\textwidth} }
\vskip -0.25 true in
\caption {Surface brightness map of a rippled slab with $kH = 0.05$, $a
  = 15$, and $\theta = 30^\circ$. Displacements on the sky are scaled in
  units of $kH$ and only one quadrant is shown.  The nested contours
  correspond to brightness enhancement factors of 10 and 15 relative to the
  background; contrast can be increased by decreasing $kH$ and increasing
  $a$. Compare the morphology to Figures 1 and 2 in Tanvir et al.~(1997)
  and Figure 8 in Hester et al.~(1995).}
\label{fig:CrabKH3}
\end{figure*}

Narrower structures with higher brightness contrasts are possible if the
emissivity of the equatorial zone is non-uniform. In particular, we might
expect the slab to be edge-brightened, since the magnetic field strength
will decline towards the center of the slab (since it reverses sign in this
region), and the acceleration of relativistic electrons might also be
greatest in the regions of strong shear and large field gradients.  If the
emissivity is confined to a fraction $\xi$ of the slab's thickness, then
the maximum surface brightness contrasts will be enhanced by an additional
factor of $\sim \xi^{-1}$. We would also expect to see a doubling of the
wisp structures, for which there may be evidence in the WFPC2 image
presented by Hester et al.~(1995; Fig.~8).

\section{DISCUSSION}

I have proposed that the moving wisps near the center of the Crab Nebula
can be interpreted as long-wavelength Kelvin-Helmholtz instabilities in the
equatorial zone of the shocked pulsar wind.  This interpretation explains
why the wisp pattern persistently moves outward from the pulsar, as
revealed in recent observations (e.g., Tanvir et al. 1997), instead of
remaining stationary or oscillating in position.  It also explains the
apparent confinement of the wisps to a plane oriented perpendicular to the
inferred pulsar rotation axis.  The observed pattern speeds of $\ltorder c/
3$ are consistent with equatorial outflow velocities of $\gtorder 0.6 c$,
shearing against a high-latitude flow with speeds of much less than $c/3$.
If the speed of the high-latitude flow is taken into account, constraints
on the equatorial flow speed are relaxed somewhat.

The wisps' appearance as narrow, high-intensity arcs of emission need not
imply that they are dissipative structures. They could simply be geometric
artifacts resulting from the projection of our line of sight through the
rippled, radiating slab. Such an interpretation is workable provided that
the dominant K-H modes have sufficiently long wavelengths ($k_R H \ltorder
0.05$) and large amplitudes. At such large amplitudes, nonlinear effects
(secondary instabilities, etc.) may be important, and the actual shapes of
the ripples are likely to be far from the simple sine wave I assumed in
\S~3 for illustrative purposes. Dissipative processes (e.g., second-order
Fermi acceleration, reconnection) could be associated with the complex
nonlinear flow so one cannot rule out the possibility that localized
dissipation contributes to the appearance of the wisps. To reconcile the
observed spacing of the wisps with the geometric model for their appearance
would require the sheet to be thin indeed, $H/r \ltorder 0.01$, where $r$
is the radial distance of the wisps from the pulsar.  Note that our model
for the Kelvin-Helmholtz instability is simplified by assuming a ``top-hat"
velocity profile rather than a smooth gradient, but this detail should not
be important for such small values of $kH$.

A more substantial worry is the stringent requirement on the slab's
emissivity compared to that of its surroundings.  To avoid the wisps' being
drowned out by a bright background, the equatorial zone's intrinsic
emissivity would have to be many times larger than that of the
high-latitude regions, since the latter occupy so much more volume. At
present I have no explanation for why this should be the case, although one
might suspect that particle acceleration is especially efficient near the
neutral sheet, where the magnetic fields are highly dynamical and
reconnection may be occurring.

I have tentatively identified the equatorial zone with the ``neutral sheet"
across which the direction of the toroidal magnetic field reverses. The
physical thickness of the shear layer ($H$) is presumably determined by
physical processes occurring in the field-reversal region. Why should there
be shear between the equatorial zone and the high-latitude regions of the
wind, especially downstream of the main wind shock where one might expect
both regions to be decelerated to $c / 3$?  One possibility is that the
equatorial zone has a higher ram pressure upstream of the shock, because of
uncertain physical processes near the base of the pulsar wind. (Note that
the ram pressure has components due to both the particle flux and the
Poynting flux.) When the high-latitude wind shocks it is slowed down to
$\sim c/ 3$ and subsequently decelerates further, but the plasma in the
equatorial zone punches through the shock radius with little change in
velocity. The equatorial zone then comes into pressure equilibrium with the
high-latitude shocked wind through a system of oblique shocks, which slows
it somewhat but still can leave substantial shear.

It is also possible for the equatorial wind zone to accelerate relative to
the high-latitude wind, even if both regions initially co-move at $\sim
c/3$ after passing through a strong shock upstream of the wisps.  The
reason is that the high-latitude wind is more highly magnetized than the
plasma in the equatorial zone where, effectively, the magnetic field
vanishes. If the high-latitude wind maintains a well-organized toroidal
field downstream of the shock, then the {\it total} pressure (gas +
magnetic) in the decelerating high-latitude wind will actually {\it
  decrease} with radius, due to the effects of magnetic tension (Begelman
\& Li 1992).  The equatorial zone will try to maintain pressure balance
with the high-latitude wind but, because its magnetization is much smaller,
it will respond to the pressure gradient by accelerating. The bulk Lorentz
factor in adiabatic flow of an unmagnetized relativistic fluid varies as
$p_{\rm tot}^{-1/4}$. Thus, in order for the equatorial zone to accelerate
from $u_0 = 1/3$ to $u = 0.5$, 0.6, or 0.7, the pressure would have to drop
to 0.71, 0.52, or 0.33 of its value immediately outside the shock. Such
pressure drops are possible provided that the magnetic pressure in the
high-latitude wind is not much smaller than the gas pressure (Begelman \&
Li 1992).

A near-equipartition magnetic field in the wisp region is seriously at odds
with widely accepted theoretical models of the Crab Nebula (Rees \& Gunn
1974; Kennel \& Coroniti 1984; Emmering \& Chevalier 1987), which imply
that the magnetic pressure in the wisp region is only a couple of percent
of the gas pressure.  (This is consistent with a ratio of Poynting flux to
kinetic energy flux in the pre-shock wind of $\ltorder 0.003$.) However,
this inference depends critically on the assumption that an axisymmetric
toroidal field structure is maintained throughout the nebula. Begelman
(1998) has pointed out that such a field geometry is highly unstable, and
that the rearrangement of the field will seriously weaken any theoretical
upper limits on the wind's magnetization.  One cannot rule out the
possibility that the shocked wind is sufficiently magnetized to sustain a
modest pressure drop, before reaching radii beyond which instabilities wash
out further pressure gradients.

Because of the many uncertainties regarding the structure of the Crab
pulsar wind and its nebular environment, it may prove difficult to devise
definitive observational tests of the model presented in this paper.
However, if the model is correct qualitatively, we expect the following
observational trends:

\begin{enumerate}
  
\item An inverse relationship between wisp spacing and pattern speed. If
  the wisps represent the nonlinear development of the Kelvin-Helmholtz
  instability, we should not be surprised to find fluctuations in the wisp
  spacing and pattern speed, even if the wind and main shock system are
  relatively steady.  Since the wavelength and phase velocity are inversely
  correlated for long-wavelength K-H modes (with $v_p \proptwid
  \lambda^{-1/2}$ asymptotically), there should be a qualitatively similar
  relationship between wisp spacing and pattern speed.  This is probably
  the most distinctive prediction of this model, compared to others.  (For
  example, the pattern speed of magnetosonic waves should be independent of
  wavelength.)
  
\item Possible lack of evidence for localized dissipation in the wisps.  If
  the wisps were dissipative structures, one might expect to see evidence
  of fresh particle acceleration (e.g., flattening of the synchrotron
  spectrum) associated with the brightest features. If the wisps are
  instead artifacts of the viewing geometry, then the spectrum should not
  be correlated with surface brightness. (Note that this test may be
  complicated by systematic variations of the emitted spectrum with height
  within the equatorial layer.)  This characteristic is shared by other
  models in which the wisps arise from adiabatic compression of
  relativistic particles, e.g., in the model of Gallant \& Arons (1994) and
  the synchro-thermal instability of Hester et al.~(1995).
  
\item Doppler asymmetry of the approaching and receding wisps. A careful
  analysis of the brightness asymmetry should be consistent with bulk flow
  velocities somewhat {\it larger} than the pattern speeds.  It will be
  important to take into account light travel time effects as well as
  aberration of the observed ripple pattern.  This should be a general
  characteristic of any model in which the outward pattern speed is slower
  than the underlying fluid velocity, i.e., it should apply in propagating
  wave models provided that the waves propagate {\it inward} relative to
  the fluid.

\end{enumerate}

\acknowledgments I am grateful to Martin Rees and Ellen Zweibel for helpful
discussions, and to the anonymous referee for constructive criticism.  This
work was supported in part by National Science Foundation grant
AST-9529175.

\end{document}